# Structural and optical changes induced by incorporation of antimony into InAs/GaAs(001) quantum dots

A. G. Taboada,1 A. M. Sánchez,2 A. M. Beltrán,3 M. Bozkurt,4 D. Alonso-Álvarez,1 B. Alén,1 A. Rivera,1 J. M. Ripalda,1 J. M. Llorens,1 J. Martín-Sánchez,1 Y. González,1 J. M. Ulloa,4 J. M. García,1,5 S. I. Molina,3 and P. M. Koenraad4

1 Instituto de Microelectrónica de Madrid, CNM (CSIC), C/Isaac Newton 8, PTM, 28760 Tres Cantos, Madrid, Spain
2 Department of Physics, University of Warwick, Coventry CV4 7AL, United Kingdom
3 Departamento de Ciencia de los Materiales e I. M. y Q. I., Facultad de Ciencias, Universidad de Cádiz, Campus Río San Pedro s/n, Puerto Real, Cádiz 11510, Spain
4 COBRA Inter-University Research Institute, Department of Applied Physics, Eindhoven University of Technology, P.O. Box 513, NL-5600 MB Eindhoven, The Netherlands
5 Department of Physics, Columbia University, New York, NY 10027, USA

We present experimental evidence of Sb incorporation inside InAs/GaA(001) quantum dots exposed to an antimony flux immediately before capping with GaAs. The Sb composition profile inside the nanostructures as measured by cross-sectional scanning tunneling and electron transmission microscopies show two differentiated regions within the quantum dots, with an Sb rich alloy at the tip of the quantum dots. Atomic force microscopy and transmission electron microscopy micrographs show increased quantum-dot height with Sb flux exposure. The evolution of the reflection high-energy electron-diffraction pattern suggests that the increased height is due to changes in the quantum-dot capping process related to the presence of segregated Sb atoms. These structural and compositional changes result in a shift of the room-temperature photoluminescence emission from 1.26 to 1.36 µm accompanied by an order of magnitude increase in the room-temperature quantum-dot luminescence intensity.

## I. INTRODUCTION

A large research effort has been focused on the fabrication of optoelectronic devices based on self-assembled semiconductor nanostructures.[1–3] For long-wavelength applications, the InAs/GaAs(001) quantum-dot (QD) emission can be redshifted by using InGaAs metamorphic layers,[4,5] or by stacking multiple layers of coupled QD,[6] but quite generally the luminescence intensity tends to decrease at longer wavelengths due to the presence of a higher concentration of atoms with a larger covalent radius, resulting in a higher accumulated stress, and consequently higher defect densities. The antimonide-based nanostructures in a GaAs matrix have been much less studied than the case of InAs QD. Different approaches have been reported: Timm et al.[7] reported on the changes induced by Sb deposition before and during QD nucleation. Similarly, Guimard et al.[8] have reported 1.3 µm emission using InAsSb QD surfactant mediated growth combined with InGaAs strain reducing layers.[9] This initial work, based on metal-organic vapor-phase epitaxy, suggested that Sb acts mainly as a surfactant during InAs QD nucleation, without significant incorporation in the QD. Ripalda et al.[10] also reported on the optical properties of InAs QDs capped with GaSb (with simultaneous codeposition of solid source Ga and Sb). [11] Several research groups have studied the use of GaAsSb strain reducing layers on top of InAs QDs.[12–16] Because of the strong segregation of Sb during GaAsSb growth on GaAs, the first monolayer (ML) on top of the QDs have a very low Sb content.[17] Here we report on the changes induced by solid source Sb deposition (without codeposition of Ga or As), after InAs QD formation and immediately before QD capping with GaAs. Because Sb is in this case directly in contact with the QDs, the composition profile and resulting electronic structure is expected to be significantly different from previous works based on the growth of GaAsSb strain reducing layers. Sb incorporation inside an InGaAs alloy is generally not expected due to the reported existence of a thermodynamic spinodal decomposition instability of the InGaAsSb quaternary.18 The resulting miscibility gap covers most of the quaternary composition range. Such spinodal decomposition effect of the InGaAsSb quaternary is expected to induce the nucleation of domain walls between regions of either InAs or GaSb composition. But in systems such as QDs, where local strain due to lattice mismatch plays an important role in the energy balance of the system, the formation of the quaternary might be energetically favorable. The Sb composition profile we have measured shows two differentiated regions within the quantum dots, whereas the approaches used by other authors either lead to no Sb incorporation inside the quantum dots,14 or to a homogeneous Sb concentration profile.11 The demonstration of quantum nanostructures with two separate regions of different composition opens the possibility to separately tune the confinement for the hole and the electron wave functions.

With the aim of reaching a better understanding of the structural, electronic, and compositional changes induced by Sb exposure, the samples were characterized by crosssectional scanning tunneling microscopy (X-STM), atomic force microscopy (AFM), in situ accumulated stress measurements, in situ reflection high-energy electron-diffraction (RHEED), photoluminescence (PL) spectroscopy, and transmission electron microscopy (TEM). The growth parameters corresponding to optimum optical characteristics for the QD ensemble are reported. We have found an order of magnitude improvement of the room-temperature (RT) luminescence intensity emission at 1.3 μm and a narrow energy dispersion of 18.8 meV full width at half maximum.

## II. METHOD.

Samples were grown by solid source molecular-beam epitaxy (MBE) on semi-insulating GaAs (001) substrates. QD nucleation was observed by RHEED after deposition of 1.65 ML of In at 510 °C substrate temperature and at 0.02 ML/s In growth rate under As4 pressure. Just after the nucleation of QDs is detected, the substrate temperature is decreased to 440 °C while In deposition continues up to a thickness of 2.2 ML. Immediately after InAs QD growth and before GaAs capping, the QDs were exposed to an Sb flux from a valved cracker Sb cell. The Sb exposure step had a duration of 10 s at beam equivalent pressures ranging from 2.0 10−7 to 3.110−6 mbar. In control samples, the Sb exposure step was substituted by a 10 s As4 exposure step at a beam equivalent pressure of 2.010−6 mbar. In our MBE system, this value corresponds to an As incorporation rate of 1.5 ML/s, calibrated by recording RHEED oscillations during group V limited GaAs growth. The GaAs capping temperature was 460 °C for the first 20 nm and then the temperature was ramped at 0.4 °C/s up to 580 °C during a 5 min growth interruption. The capping was completed at this temperature up to a total thickness of 105 nm. Samples for AFM imaging were grown in the same conditions, but the sample growth was interrupted, and the sample temperature quenched, immediately after the 10 s As/Sb exposure. For structural characterization by X-STM and TEM, a stack with four QD layers was also grown. The thickness of the GaAs spacer layers was 70 nm. The first layer is the reference InAs QD layer and the other layers were exposed to Sb during 10 s at the same Sb beam equivalent pressures used previously in single-layer QD samples: $2.0 \times 10^{-7}$, $8.0 \times 10^{-7}$, and $3.1 \times 10^{-6}$ mbar.

X-STM images were obtained at room temperature in an ultrahigh vacuum chamber ($p<10^{-11}$ mbar) on the in situ cleaved (110) surface plane. Polycrystalline tungsten tips prepared by electrochemical etching were used. All the images shown in this paper were recorded at voltages higher than 2.0 V, which is enough to suppress the electronic contrast.19 Specimens for cross-sectional TEM were prepared by mechanical grinding to below 20 μm and mechanical polishing using 1 μm diamond suspension on a soft nap pad.20 The specimens were then ion milled using Ar+ ions at 4.5 kV and a beam incidence angle of 3° to electron transparency. A final low-energy step at 2 kV was employed to minimize amorphous surface layers. Conventional TEM was carried out on JEOL 1200EX and Jeol 2000FX microscopes operating at 120 kV and 200 kV, respectively, close to the ⟨110⟩ GaAs zone axis. Images were recorded in dark-field two beam conditions, using the 002 reflection. Highly sensitive in situ optical measurements of the substrate curvature were used to measure the accumulated stress ($\sum \sigma$) evolution during heteroepitaxial growth.21,22 The samples are cantilever shaped (typically 5x20 mm$^2$) GaAs substrates with a thickness of 100 μm, mechanically clamped on one end, while the other end is free to move. The angle of reflection of a laser beam was compared at the clamped end and the free end of the substrate. The $\sum \sigma$ evolution was measured for two different Sb exposures: $3.6 \times 10^{-7}$ and $3.4 \times 10^{-6}$ mbar. In control experiments Sb4 was substituted by As4. PL was excited with a frequency-doubled Nd: yttrium aluminum garnet laser $\lambda_{exc}=532$ nm focused to a spot diameter of approximately 150 μm and attenuated to 10 mW. The corresponding PL spectrum was dispersed with a 0.3 m focal length spectrometer and detected using a Peltier cooled InGaAs photodiode array.

## III. RESULTS

### A. Ensemble characterization: AFM, PL, and RHEED

A comparison of 2x2 μm$^2$ AFM micrographs of QD ensembles (a) without and (b) with Sb exposure is shown in Fig. 1. Even moderate Sb exposures completely change the density and size distribution of the QD ensemble. To quantify these changes, we have carried out an AFM statistical study of QD surface density vs height for different Sb exposures. The resulting histograms, with a 1 nm height binning, are shown in Figs. 2(a)–2(f). The ensemble without Sb shows a broad monomodal distribution centered at 8 nm with a standard deviation of w=1.45 nm. At intermediate Sb exposures [Fig. 2(c)], there is a splitting of the size distribution into two subensembles, centered at 12.0 nm and 4.5 nm with a standard deviation of w=0.79 nm and w=0.99 nm respectively. Increasing PSb to $8.0 \times 10^{-7}$ mbar, the smaller QDs disappear leading to an optimal size distribution with a narrow monomodal distribution (w=0.78 nm) centered at 13.0 nm [Fig. 2(d)]. At higher Sb doses, the AFM analysis reveals that the dot height saturates and then starts to decrease while the QD surface density remains constant at around $8 \times 10^9$ cm$^{-2}$ and the size distribution broadens. The QD surface density starts at $\delta=1.5 \times 10^{10}$ cm$^{-2}$ for the InAs sample, reaches a minimum value of $\delta=5 \times 10^9$ cm$^{-2}$ at $P_{Sb}=8.0 \times 10^{-7}$ mbar, and then rises again to $\delta=8 \times 10^9$ cm$^{-2}$ for $P_{Sb}=3.1 \times 10^{-6}$ mbar. In the reference sample without Sb exposure, the surface density of QDs with height >17 nm is $2.6 \times 10^8$ QDs/cm$^2$.

These oversized QDs are typical of InAs QD ensembles, and are known to be a cause of nonradiative recombination, as they are plastically relaxed.23 Remarkably, these oversized QDs have not been detected in our QD ensembles exposed to moderate Sb pressures [Fig. 1(a)] compared to Fig. 1(b). One of the factors affecting quantum dot size is the In sublimation rate at the growth temperature. The formation of InSb during Sb exposure is limited by the very large increase in the strain induced by Sb incorporation. This strain energy barrier to the formation of In-Sb bonds can be expected to be much smaller at the plastically relaxed QDs than in the elastically strained QDs. At the growth temperature used in our work, the evaporation rate of InSb is significantly higher than the evaporation rate of InAs, limiting the maximum size of the plastically relaxed QDs. Room-temperature PL spectra of capped QD samples are shown in Figs. 2(g)–2(l). In each spectrum, the peak at shorter wavelength is the first excited state. The exposure of QD to Sb results in an emission redshift of up to 73 meV. According to electronic-structure calculations of InAs QD as a function of QD height published in the literature,24,25 the expected redshift of the luminescence as the QD height increases from 7 to 12 nm is 70 meV. Our quantum dots are not expected to be composed of pure InAs but this value gives an order of magnitude estimate of the effect of QD height on the electronic structure.

The intensity of the luminescence decreases at Sb pressures above $1.6 \times 10^{-6}$ mbar and also shows a local minimum at $P_{Sb}=4.010-7$ mbar [Fig. 2(i)]. This minimum coincides with the appearance of a bimodal distribution and a minimum density of the larger QDs as observed by AFM [Fig. 2(c)]. Only the larger QDs are expected to significantly contribute to the PL at room temperature due to thermal escape of excitons out of the smaller QDs. Figure 3 shows the time evolution of the intensity of the 002 transmission diffraction spot that is used to detect the presence of quantum dots on the surface during growth. Immediately after the Sb flux exposure starts, the intensity of the transmission spots decreases abruptly compared with the case in which the QD are exposed to an As4 flux. We relate this behavior to the formation of a floating Sb layer without crystalline long-range order on top of the surface. This amorphouslike layer reduces the intensity of the RHEED pattern. Remarkably, when QD are exposed to Sb, the transmission spots remain visible for a longer time during the capping process. This is in good agreement with the works by Liu et al.14 and Ulloa et al.26 which reported a better preservation of QD size when Sb is present in the capping layer. This is also consistent with the longer emission wavelength in comparison with InAs QDs not exposed to a Sb flux.

**B. Structure of capped QD.**

**1. Transmission electron microscopy.**

The sample with four stacked QD layers was characterized by TEM in order to examine the variations in QD size after capping with GaAs. TEM results show good quality material with a relatively low defect density ($<10^7$ cm$^{-2}$). Figure 4(a) shows a dark-field 002 TEM overview of the four layer sample used for cross-sectional structural characterization. The bottom layer corresponds to an InAs QD reference layer without Sb flux. The three upper layers have increasing Sb exposures of $P_{Sb}=2.0 \times 10^{-7}$ mbar, $P_{Sb}=8.0 \times 10^{-7}$ mbar, and $P_{Sb}=3.1 \times 10^{-6}$ mbar from bottom to top. QDs exposed to the highest Sb pressures present two bright layers at the top and bottom of the QD separated by a dark layer in the middle [Fig. 4(c)]. The QD size statistics obtained from TEM and AFM image analysis are summarized in Fig. 5. The contribution of strain fields to the contrast in DF002 TEM images extends beyond the boundaries of the quantum dots, leading to a systematic overestimate of the size of lattice mismatched nanostructures by measuring the height of the observed DF002 contrast. TEM measurements show that, after Sb exposure, the base diameter of the capped QDs remains unchanged within the uncertainty of our data, while the QD height increases with Sb exposure, almost independently of PSb. Such an increase is consistent with the redshift of the photoluminescence. The PL redshift increases monotonically with PSb while the QD average height as measured by AFM reaches a maximum at an intermediate value of PSb. The capped QDs height remains almost constant when PSb is further increased. The changes in the PL at the higher Sb exposures must be then related to changes in the composition of the QDs or the matrix around the QDs. The contrast observed in TEM images of Fig. 4 has its origin in the compositional sensitivity of the 002 reflection. Using the kinematical diffraction theory,27,28 the intensity of the diffracted 002 electron beam as a function of composition and the atomic scattering factors f, can be approximated by

$$I \propto [f_{In}x - f_{Sb}y + f_{Ga}(1-x) - f_{As}(1-y)]^2. \quad (1)$$

Figure 6 shows the calculated 002 diffracted intensity as a function of composition for the $In_xGa_{1-x}As_{1-y}Sb_y$ quaternary normalized to the intensity for x=y=0. The dark area in Fig. 6 corresponds to alloy compositions with y~x−0.22. We therefore interpret the dark region in the middle of the QDs in Fig. 4(c) as a region with an alloy composition roughly meeting the y~x−0.22 condition, and therefore as evidence of the presence of Sb atoms inside the QDs, as the In content inside QDs is typically well above 22%.29 An alternative explanation for our TEM images would be an exceptionally low In concentration inside the QD but that would be incompatible with the observed emission wavelength in the PL spectra. As can be seen in Fig. 4(d), reference InAs QDs without Sb exposure do not show the dark intermediate layer seen in Figs. 4(b) and 4(c).

**2. Cross-sectional scanning tunnel microscopy**

With the aim of obtaining quantitative composition and lattice parameter profiles we have also conducted crosssectional STM measurements on (110) surfaces cleaved in ultrahigh-vacuum conditions. In Figs. 7(a) and 7(b) we compare filled-states topography images of individual QDs exposed to PAs=2.010−6 mbar and PSb=8.010−7 mbar, respectively. The lines perpendicular to the [001] growth direction represent the III-V zigzag chains at the zinc-blende (110) surface. To obtain a count of Sb atoms inside the QDs, we have analyzed the X-STM filled-state images after Fourier filtering the low-frequency components of constant tunneling current images to remove the effects of outward relaxation. Examples of such images are shown in Fig. 7. Because of differences in the local electronic structure, and the larger covalent radius of Sb compared to As, the tunneling tip retracts slightly above each Sb atom to keep the tunneling current constant. We have traced profiles along the growth direction in order to measure the STM tip retraction above each group V atom relative to the trenches immediately above and below. We have only counted as Sb those atoms associated with a tip retraction larger than the maximum tip retraction observed in the InAs reference sample. The resulting composition values are therefore likely to be lower than the true Sb composition and only reflect a lower limit to the Sb content. The uncertainty in these data is a consequence of the random distribution of buried In atoms and Sb atoms in the $Ga_xIn_{1-x}As_ySb_{1-y}$ quaternary alloy causing fluctuations in the surface morphology that can only be partially cancelled by Fourier high-pass filtering. The resulting lower limits for the Sb concentration that we measure inside the QD are collected in Table I. Separate values are given for the upper and lower halves of the QD. The Sb content is significantly higher in the top half, suggesting a possible explanation for the contrast observed in the TEM images.

Atomic plane spacing profiles taken along the growth direction through the center of cleaved QD are presented in Fig. 8. They have been obtained by averaging the spacing between consecutive atom rows in several line profiles. The lattice parameter profiles measured for the InAs QD control layer reproduce the results reported in the literature,29 corresponding to an In concentration variation from x=0.8 at the base to x=1.0 at the top of $In_xGa_{1-x}As$/GaAs QD. The lattice parameter at the apex of the InAs QD exposed to the highest Sb pressure is 0.71 nm, clearly larger than the atomic plane spacing predicted and measured for InAs QD value, and in coincidence with the atomic plane spacing measured by Timm et al.30 in GaSb QD.

| $P_{Sb}$ (mbar) | QD bottom half Sb concentration (%) | QD top half Sb concentration (%) |
|---|---|---|
| $2.0 \times 10^{-7}$ | $\geq 8.7 \pm 2.6$ | $\geq 14.0 \pm 4.4$ |
| $8.0 \times 10^{-7}$ | $\geq 5.4 \pm 2.1$ | $\geq 11.6 \pm 3.5$ |
| $3.1 \times 10^{-6}$ | $\geq 9.8 \pm 5.3$ | $\geq 18.2 \pm 8.6$ |

TABLE I. Sb concentration of the lower and top half of InAs QD exposed to different Sb pressures.

The exposure to Sb also changes the wetting layer (WL) composition. Figure 9 shows the measured X-STM outward relaxation profiles of the reference InAs WL and the one exposed to the highest Sb pressure. Figure 9 also plots the lower limit of the WL Sb content as measured by atom counting from X-STM images. The outward relaxation of the cleaved surface is related to the accumulated stress during heteroepitaxial growth and is therefore also related to the alloy composition. In our case, we observe that the integrated outward relaxation of the cleaved surface is somewhat higher in the WL exposed to Sb. For the InAs sample, the deformation profile suggests an exponentially decaying In content in the growth direction that is characteristic of slow incorporation of In from a segregated layer floating on the growth front. A similar exponential composition profile due to surface segregation is known to occur for Sb during heteroepitaxial growth on GaAs.17,30 The strain relaxation profile at high Sb exposures suggests the presence of a double peak structure with a small feature 4 nm above the WL.

## 3. In situ characterization: Stress accumulation

We have measured the accumulated stress during growth by measuring the differential deflection of a laser beam impinging on a cantilever-shaped sample. As the main driving force of QD formation is the strain caused by lattice mismatch, in situ stress measurements provide valuable information about the QD growth process.31 The measured cantilever shaped substrate curvature is related to the accumulated stress in the epitaxial layer by

$$\frac{-Y_s h_s^2}{6R(1-\nu_s)} = \Sigma\sigma + \tau, \qquad (2)$$

where R is the curvature radius of the cantilever, $h_s$ is the substrate thickness, $\Sigma\sigma$ is the accumulated stress in the epitaxial layer, $\tau$ is the surface tension, $Y_s$ is the Young's modulus of the substrate, and $\nu_s$ is the Poisson ratio of the substrate. Figure 10(a) shows the accumulated stress evolution along [110] direction during formation and capping of InAs QD exposed to As and two different Sb pressures. The evolution of the accumulated stress during capping of selfassembled InAs QDs has been studied before by García et al.31 The change in slope during In deposition is due to the nucleation of quantum dots. The rapid raise of the accumulated stress during capping with GaAs is due to the incorporation to the bulk of the crystal of In and Sb atoms that were not effectively contributing to the stress, as they were either segregated over the growth front as a surfactant layer, or were at the tips of the QDs, where the lack of constrains allows for elastic relaxation. The accumulated stress $\Sigma\sigma$ is given by

$$\Sigma\sigma = \int \sigma(z)dz = \int M(z)\varepsilon(z)dz, \quad (3)$$

where $\varepsilon = -0.07164(x+1.09y)$ is the strain introduced by a quaternary $In_xGa_{1-x}As_{1-y}Sb_y$ alloy epitaxially grown on GaAs,32 and M is the biaxial modulus given (in $10^{10}$ N/m$^2$) by

$$M = c_{11} + c_{12} - 2c_{12}^2/c_{11} = 12.39 - 4.45x - 3.19y + 1.62xy. \quad (4)$$

The derivative of $\Sigma\sigma(z)$ is the stress $\sigma(z)$ as a function of z and is related to the alloy composition by

$$\sigma = ax + by + cxy + dx^2 + ey^2 + fx^2y + gxy^2, \quad (5)$$

where the a, b, c, d, e, f, and g factors are −8.87, −9.67, 5.76, 3.19, 2.49, −1.16, and −1.27 (in $10^9$ N/m$^2$), respectively.33 Thus, the effect of Sb incorporation on the accumulated stress is only slightly larger than the effect of In, and for x, y≪1 the accumulated stress is approximately proportional to the sum of the In and Sb content in the grown layers. During the first 8 ML of capping, all the samples, with or without Sb, approximately accumulate the same stress, 2.4 N/m. The effects of segregated In on the incorporation rate of Sb, and of segregated Sb on the incorporation rate of In, have been studied by Haxha et al.17 and Sanchez et al.34 Applying Eq. (5), the stress accumulated during the first 8 ML of capping is equivalent to 8 ML of an $In_xGa_{1-x}As_{1-y}Sb_y$ alloy approximately meeting the condition 0.125≥x+y≥0.112. For the sample without Sb, x=0.125 and y=0.

The total amount of Sb incorporated can be estimated by means of Eq. (5) and the $\Sigma\sigma$ curves shown in Fig. 10(a). We assume here that the total amount of incorporated In is the same for all three samples when the capping is finished (this is equivalent to neglecting the possible effect of Sb on the desorption rate of In), and neglect surface tension changes, which as discussed below, are an order of magnitude smaller than the accumulated stress changes observed during capping. For the samples exposed to 3.6x10$^{-7}$ mbar and 3.4x10$^{-6}$ mbar of Sb partial pressure, this procedure yields a Sb incorporation of 0.4 ML and 0.6 ML, respectively. The total Sb exposure ranges from 0.4 ML to 3.8 ML (in terms of group V limited growth rate) corresponding, respectively, to Sb beam equivalent pressures of 3.6x10$^{-7}$ –3.4x10$^{-6}$ mbar during the Sb exposure step. Surprisingly, increasing one order of magnitude the Sb pressure does not significantly change the amount of incorporated Sb. The incorporation of Sb saturates at 0.6 ML and is then almost independent on the Sb beam pressure.

We have also measured the $\Sigma\sigma$ in two different crystallographic directions: [110] and [1-10] [Fig. 10(b)]. The curve corresponding to the [110] azimuth in Fig. 10(b) is an independent measurement in the same experimental conditions as the curve corresponding to $P_{Sb}$=3.6x10$^{-7}$ mbar in Fig. 10(a). When the surface is exposed to Sb, an azimuth-dependent feature appears in the accumulated stress. Such anisotropic changes in the accumulated stress and surface tension can be attributed either to surface reconstruction changes,35 to the effect of surfactants, or to the formation of anisotropic nanostructures such as quantum wires.36 The latter mechanism seems less plausible, since the accumulated stress change introduced by Sb exposure (0.5 N/m) is comparable with the stress introduced by reconstruction changes in GaAs,37 and no clear morphological anisotropy has been found in AFM images.

The accumulated stress measurements reveal that the main Sb incorporation occurs during GaAs capping. To separate the contributions to the accumulated stress produced by the QD and by the WL, we have grown a sample with WL but without QD, by depositing 1 ML of In on GaAs and exposing the growth front during 10 s at $P_{Sb}=3.6\times10^{-7}$ mbar. The corresponding accumulated stress curves are presented in Fig. 10(b). The accumulated stress evolution corresponding to the sample without QD and the samples with QD are remarkably similar except during the Sb exposure step. The well-known surfactant effect of Sb on III-V semiconductor surfaces implies a reduction in surface tension.38 As the accumulated compressive stress is negative, such a reduction in surface tension might appear as an increase in the accumulated compressive stress. This is the most likely explanation for the almost immediate change in the sample curvature observed as the Sb flux is turned on.

The sample without QD almost immediately accumulates 0.5 N/m and then remains flat through the Sb exposure step. In contrast, the samples with QD present a more complex time dependence of the accumulated stress, presumably due to the interaction of Sb with the QD. Remarkably, the stress accumulated during capping is the same (5 N/m) for both types of samples, with and without QD.

**IV. DISCUSSION.**

The changes in QD density and size before capping can be related with the enhancement of the ripening effect in the presence of an Sb flux reported by Pötsche et al.38,39 During this ripening process, some QD grow at the expense of others, changing the size distribution. During the Sb exposure step, almost one half of the supplied In is still segregated on the surface and not yet incorporated into the crystal as evidenced by the stress accumulation measurements.31 After Sb exposure the density of oversized clusters decreases. This effect is related to the formation of In-Sb bonds during exposure to Sb. The formation of InSb during this Sb exposure is limited by a very large increase in the strain in the pseudomorphic quantum dots. Oversized clusters are known to be metamorphic, so in this case there is a much smaller strain energy barrier to the formation of In-Sb bonds, which are weaker than In-As bonds. At the growth temperature used in our work, the evaporation rate from the surface of InSb is significantly higher than the evaporation rate of InAs. This As-Sb exchange reaction also explains the sudden increase in the in situ accumulated stress measurements [Figs. 10(a) and 10(b), from t=−10 to 0 s]. The subsequent stress relief can be explained by the loss of some InSb from the surface at that temperature.

It has been reported that the dependence of QD energy levels on QD size fluctuations decreases as QD size increases.14,40 This effect and the narrower size distribution observed by AFM explain the observed narrowing of the luminescence peaks at moderate Sb exposures. There are two possible causes of the PL redshift and the enhancement of the luminescence intensity: an increase in the capped QD size and changes in the electronic structure caused by Sb incorporation. Both effects increase the carrier thermal activation energies, whereas the incorporation of Sb in the nanostructures also implies an upward shift of the conduction and valence bands and consequently a deeper hole confinement.41 RT luminescence is typically limited by hole thermal escape and thus a deeper hole confinement would explain the observed RT PL enhancement.42 Sb concentration profiles show incorporation of Sb in the top layers of the QD. It is well known that the lattice parameter is larger at the tip of uncapped QDs than at any other point in the crystal surface.43,44 This stretching of the lattice at the dot apex facilitates the incorporation of atoms of a bigger covalent radii, such as Sb.

The effects of Sb exposure are mainly observed during GaAs capping. The strong segregation of Sb during heteroepitaxy on GaAs is due to the larger covalent radius of Sb and also to the fact that the Sb-Sb bond is stronger than the Ga-Sb bond (2.6 eV and 1.5 eV, respectively). 45,46

## V. CONCLUSIONS.

Both X-STM and TEM images show evidence of an Sbrich region at the tip of InAs QDs exposed to Sb immediately before capping. The changes induced by Sb exposure on the structure and composition of InAs QDs are summarized in Fig. 11. Although there is Sb incorporation at the tip of the QDs exposed to Sb flux, its concentration does not show a strong dependence on the Sb beam pressure above Sb exposures equivalent to 0.4 ML (in terms of GaSb monolayers heteroepitaxially grown on GaAs), and saturates at 0.6 ML. At low Sb pressures, there is an increment of the size of the QD before capping. Under optimal conditions, InAs QDs exposed to Sb have room-temperature PL emission at 1340 nm with a full width at half maximum of 18.8 meV and with an intensity one order of magnitude higher than InAs QD control samples. For our experimental set up, the optimal Sb exposure is 10 s at $8.0 \times 10^{-7}$ mbar. The cause of such changes in the optical characteristics has been investigated with in situ characterization during growth and structural microanalysis. Both the change in the QD height and the composition changes at the tip of the QDs cause the change in the optical properties. The smaller electron affinity of antimonides shifts both the conduction and the valence band closer to the vacuum level, further increasing the thermal escape barrier for holes, which is most often the limiting factor for the room-temperature luminescence intensity of InAs QD.

ACKNOWLEDGMENTS

Support by CAM (Projects No. S-505/ENE-310, No. S-505/ESP/000200, and No. S2009ESP-150), by MEC (Project No. TEC2008-06756-C03-01), Consolider-Ingenio 2010 QOIT (Grant No. CSD2006-00019) and GENESIS (Grant No. CSD2006-00004), and by MICINN (FPI grant of AGT) is acknowledged. A.R. thanks the I3P program of CSIC for financial support. A.M.S. would like to thank the Science City Research Alliance and the HEFCE Strategic Development Fund for funding Support.

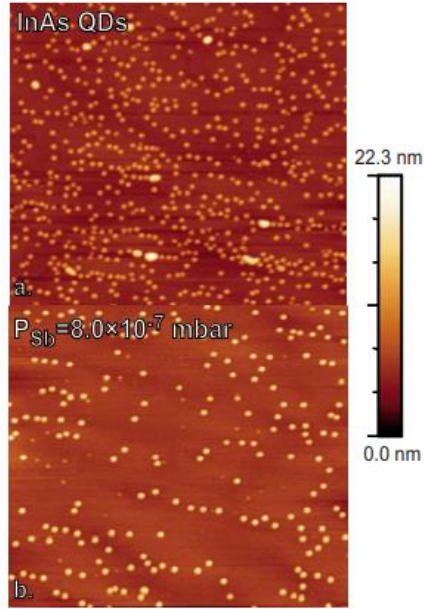

FIG. 1. 2x2 μm$^2$ AFM images of InAs QD ensembles (a) without Sb exposure and (b) after a 10 s exposure to an Sb pressure of P$_{Sb}$=8.0x10$^{-7}$ mbar.

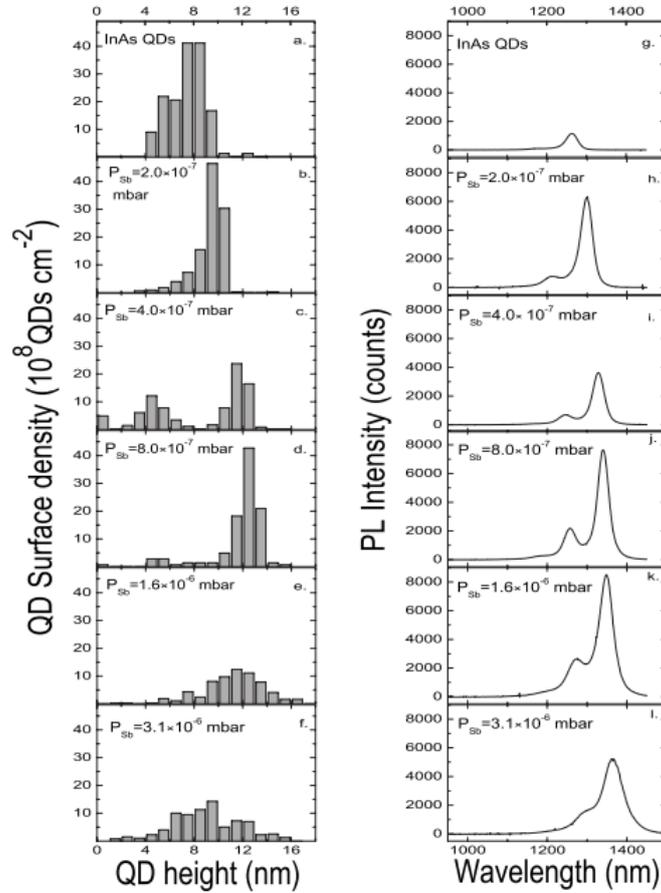

FIG. 2. [(a)–(f)] AFM statistics of QD surface density vs height with a 1 nm height binning as a function of the increasing Sb pressure at a constant exposure time of 10 s and [(g)–(l)] corresponding room-temperature PL spectra.

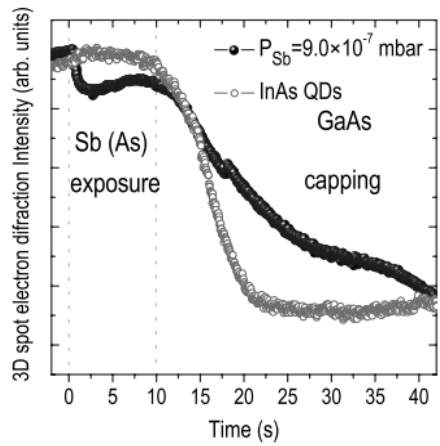

FIG. 3. QD three-dimensional (3D) RHEED pattern intensity evolution for InAs QD exposed to As and Sb followed by GaAs capping. When QD are exposed to Sb, the 3D pattern remains visible longer time during capping. The inset shows the RHEED pattern immediately before Sb.

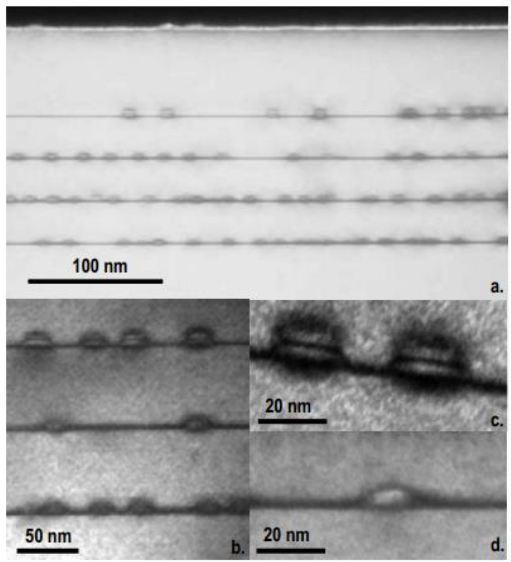

FIG. 4. Dark-field 002 TEM images of: (a) the QD four-layer stack. The bottom layer is an InAs QD reference layer, and the upper layers have increasing Sb exposures of $P_{Sb}=2.0 \times 10^{-7}$ mbar, $P_{Sb}=8.0 \times 10^{-7}$ mbar, and $P_{Sb}=3.1 \times 10^{-6}$ mbar. (b) Detail of the layers exposed to Sb. (c) Detail of the top layer QDs exposed to $P_{Sb}=3.1 \times 10^{-6}$ mbar. (d) Reference InAs QD exposed to As.

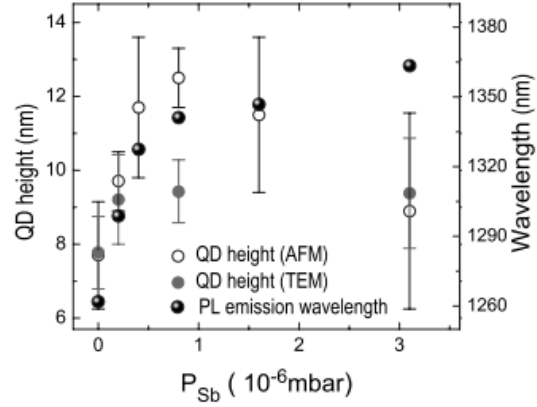

FIG. 5. InAs(Sb) QD heights (as measured by TEM and AFM) and corresponding emission wavelengths plotted against the pressure of the Sb beam during the exposure step.

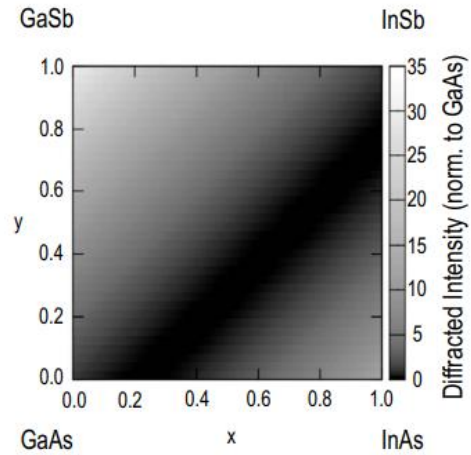

FIG. 6. Calculated diffracted 002 intensities as a function of composition in the $In_xGa_{1-x}Sb_yAs_{1-y}$ system.

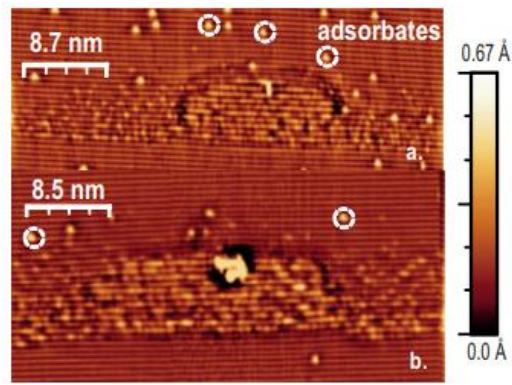

FIG. 7. Cross-sectional scanning tunneling micrographs of: (a) InAs QD exposed to $P_{As4}=2.0\times10^{-6}$ mbar. (b) InAs QD exposed to an Sb pressure $P_{Sb}=8.0\times10^{-7}$ mbar during 10 s before capping. This growth conditions optimize optical properties. The images have been flattened by removing low frequency components with a Fourier filter.

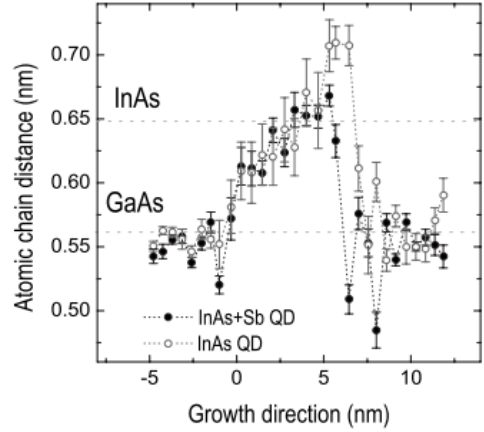

FIG. 8. Atomic plane spacing profiles taken in the growth direction for QD with and without Sb exposure at $P_{Sb}=3.1\times10^{-6}$ mbar. The origin (0 nm) is at the QD base. Error bars represent the standard error of the mean.

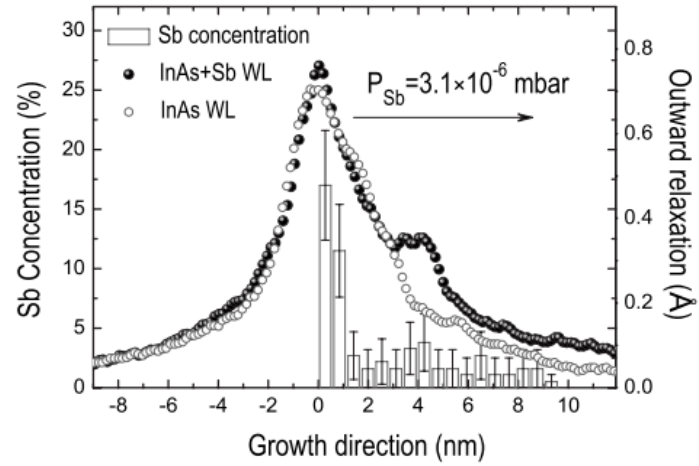

FIG. 9. Outward strain relaxation profiles of InAs reference WL and InAs WL exposed to Sb ($P_{Sb}=3.1\times10^{-6}$ mbar). The latter is compared with its Sb distribution along the growth direction [001]. Sb concentration has been estimated by atom counting. The origin (0 nm) is set in the first WL monolayer. Error bars represent the binomial proportion confidence interval at a 90% confidence level.

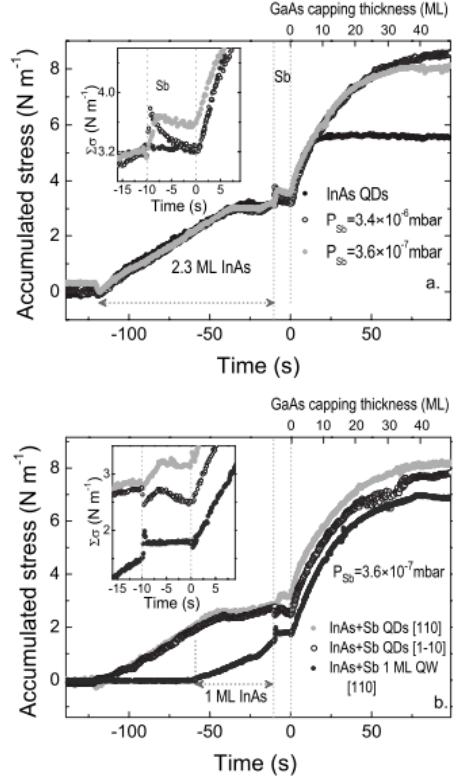

FIG. 10. Accumulated stress evolution during InAs QD formation, Sb exposure, and subsequent GaAs capping in different experiments: (a) InAs QD exposed to $P_{Sb}=3.6\text{x}10^{-7}$ mbar and $P_{Sb}=3.4\text{x}10^{-6}$ mbar, and a control experiment exposing InAs QD to $P_{As}=2.0\text{x}10^{-6}$ mbar. The inset shows the detail of the $\sum \sigma$ evolution during Sb exposure growth step for the different pressures. (b) Samples with QD and without QD, instead, with 1 ML InAs quantum well. The Sb exposure step took place at $P_{Sb}=3.6\text{x}10^{-7}$ mbar. The signal for the QD samples is shown for cantilevers cut in such a way that the long side is parallel either to the [110] or to the [1-10] crystallographic directions. The inset shows the detail of the $\sum \sigma$ evolution during the 10 s Sb exposure step in both crystallographic directions and in the quantum well.

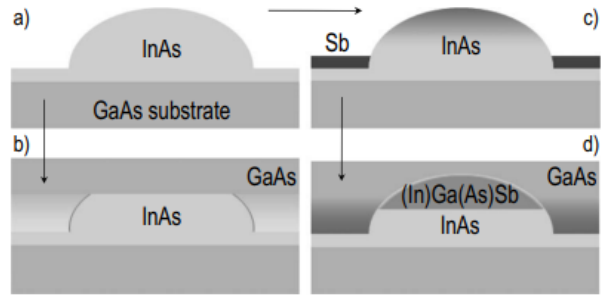

FIG. 11. Scheme of the structural changes induced by Sb exposure immediately before QD capping. (a) Initial uncapped InAs QDs. (b) Structural changes induced by GaAs capping. (c) InAs QDs exposed to Sb before capping, with segregated Sb layer, and Sb incorporation at the apex of the QDs. (d) Effects of GaAs capping after Sb exposure. The QD height is significantly higher than in case (b).